\documentclass[aps,pre,groupedaddress,showpacs,preprint]{revtex4}

\usepackage{graphicx}
\usepackage[dvips]{epsfig}
\usepackage{dcolumn}
\usepackage{subfigure}%
\usepackage{bm}
\usepackage{amssymb}
\usepackage{amsmath}

\begin{document}
\title{Field Theory of Propagating Reaction-Diffusion Fronts}

\author{C. Escudero}

\affiliation{
Departamento de F\'{\i}sica Fundamental,
Universidad Nacional de Educaci\'on a Distancia, C/ Senda del Rey 9,
28040 Madrid, Spain}

\begin{abstract}
The problem of velocity selection of reaction-diffusion fronts has been widely investigated. While the mean
field limit results are well known theoretically, there is a lack of analytic progress in those cases in
which fluctuations are to be taken into account. Here, we construct an analytic theory connecting the first
principles of the reaction-diffusion process to an effective equation of motion via field-theoretic
arguments, and we arrive at the results already confirmed by numerical simulations.

\end{abstract}

\maketitle

\section{Introduction}
\label{introduction}

Reaction-diffusion front propagation in nonequilibrium systems is a topic that has been receiving an
increasing attention recently. The numerous possible applications of the theory, as can be systems like
flames~\cite{kolmogorov},
bacterial colonies~\cite{golding}
or population genetics~\cite{fisher},
is of course one of the reasons of this recent interest.
One of the most common approaches to this problem has been the use of deterministic reaction-diffusion
equations, like the Fisher equation~\cite{fisher}.
This equation, that combines logistic growth with diffusion, is one of
the most important mathematical models in biology and ecology~\cite{murray}.
In one spatial dimension, the Fisher equation
reads
\begin{equation}
\partial_t U= D\partial_{xx}U+aU-bU^2.
\end{equation}
One can think this equation as the mean field description of a reaction-diffusion process of a single species
of random walkers $A$ undergoing the reactions of birth $A \to A+A$ at rate $a$ and annihilation
$A+A \to \emptyset$ at rate $b/2$. The analysis of this equation is straightforward. Consider the boundary
conditions $U \to b/a$ when $x \to -\infty$ and $U \to 0$ when $x \to \infty$. Thus the linearly stable
phase $b/a$ invades the linearly unstable phase $0$. Assuming a stationary front profile
U(x,t)=U(x-vt)=U(z) and shifting variables $x \to \sqrt{D/a}x$, $t \to t/a$ and $U \to (a/b)U$ we get
\begin{equation}
\label{fisher}
U''+cU'+U-U^2=0,
\end{equation}
where $c=v/\sqrt{Da}$.
The velocity of the front is controlled by its edge, this means, the region of the front that is closer to
the unstable phase $U=0$.
We can thus linearize Eq.(\ref{fisher}) around this value to get
\begin{equation}
\label{linear}
U''+cU'+U=0.
\end{equation}
The only physically acceptable solution to this equation is
\begin{equation}
\label{solution}
U(z) \sim e^{-\gamma z},
\end{equation}
and substituting Eq.(\ref{solution}) into Eq.(\ref{linear}) we get that
\begin{equation}
c=\gamma+\frac{1}{\gamma},
\end{equation}
for an arbitrary $\gamma$. It is clear that the range of velocities is thus $c \ge 2$,
and it was shown that
the minimal velocity is selected in the long time limit~\cite{aronson}.
We can thus conclude that, in this limit,
$v=2\sqrt{Da}$.

We will now show that this picture changes strongly when internal fluctuations effects, due to the finitness
and discretness of the reactants, are taken into account.

\section{The Field Theory}
\label{field}

We will consider a single species particles $A$ undergoing the reactions
$A \to A+A$ at rate $\sigma$
and $A+A \to \emptyset$ at rate $\lambda$. Further, we suppose the particles $A$ performing a random walk
in a one dimensional lattice with lattice spacing $b$.
The exact description of the problem is given by the following master equation:
\begin{equation}
\label{master}
\frac{dP(\{n_i\};t)}{dt}=\sum_i \left[ \left.\frac{dP(\{n_i\};t)}{dt} \right|_D +
\left.\frac{dP(\{n_i\};t)}{dt} \right|_\sigma +
\left.\frac{dP(\{n_i\};t)}{dt} \right|_\lambda \right],
\end{equation}
with
\begin{equation}
\left.\frac{dP(\{n_i\};t)}{dt} \right|_D=\frac{D}{b^2}\sum_{\{e\}}[(n_e+1)P(...,n_i-1,n_e+1,...;t)
-n_iP(...,n_i,n_e,...;t)],
\end{equation}
where $\{e\}$ denotes the set of nearest-neighbor sites adjacent to $i$ and $D$ is the diffusion constant,
\begin{equation}
\left.\frac{dP(\{n_i\};t)}{dt} \right|_\sigma= \sigma[(n_i-1)P(...,n_i-1,...;t)-n_iP(...,n_i,...;t)],
\end{equation}
and
\begin{equation}
\left.\frac{dP(\{n_i\};t)}{dt} \right|_\lambda =\lambda [(n_i+2)(n_i+1)P(...,n_i+2,...;t)
-n_i(n_i-1)P(...,n_i,...;t)].
\end{equation}
For simplicity we will choose an uncorrelated Poisson distribution as initial condition for our
master equation:
\begin{equation}
P(\{n_i\};t=0)=e^{-N(0)}\prod_i \frac{n_{0i}^{n_i}}{n_i!},
\end{equation}
where $N(0)=\sum_i n_{0i}$. We can map this master equation description of the system into a quantum
field-theoretic problem. This connection was first proposed by Doi~\cite{doi}, further elucidated by
Peliti~\cite{peliti} and a deep generalization of it can be found in the influencing article by Cardy and
T\"{a}uber~\cite{cardy}.
We can write this theory in terms of the second-quantized bosonic operators:
\begin{equation}
[a^\dag_i,a_j]=\delta_{ij}, \qquad [a_i,a_j]=0, \qquad [a^\dag_i,a^\dag_j]=0,
\qquad a_i \left| 0 \right>=0,
\end{equation}
whose effect is to create or to annihilate particles at the corresponding lattice site:
\begin{equation}
a^\dag_i \left|...,n_i,...\right>=\left|...,n_i+1,...\right>,
\end{equation}
\begin{equation}
a_i \left|...,n_i,...\right>=n_i\left|...,n_i-1,...\right>,
\end{equation}
where we have defined the states as:
\begin{equation}
\left|\{n_i\}\right>=\prod_i (a_i^\dag)^{n_i} \left|0 \right>.
\end{equation}
Thus we can define the time-dependent state vector as:
\begin{equation}
\label{vector}
\left|\Phi(t) \right>=\sum_{\{n_i\}}P(\{n_i\};t)\left|\{n_i\} \right>,
\end{equation}
and claim that it obeys the imaginary time Schr\"{o}dinger equation
\begin{equation}
\label{schrodinger}
\frac{d}{dt}\left|\Phi(t)\right>=-H \left|\Phi(t) \right>,
\end{equation}
with the hamiltonian
\begin{equation}
\label{hamiltonian}
H=\sum_i \left(-\frac{D}{b^2}\sum_{\{e\}}a^\dag_i(a_e-a_i)-\lambda [1-(a^\dag_i)^2]a_i^2
+\sigma [1-a^\dag_i]a^\dag_ia_i \right).
\end{equation}
Note that we recover Eq.(\ref{master}) if we substitute Eq.(\ref{vector}) and Eq.(\ref{hamiltonian}) in
Eq.(\ref{schrodinger}). The time dependent expectation value of an observable $O$ is given by:
\begin{equation}
\label{expectation}
\left< O(t) \right>= \sum_{\{n_i\}} O(\{n_i\})P(\{n_i\};t).
\end{equation}
To compute this quantity in the field-theoretic formalism we need to introduce the Glauber state:
\begin{equation}
\left< S \right| = \left< 0 \right|\prod_i e^{a_i}, \qquad \left< S | 0 \right>=0.
\end{equation}
Note that this state is a left eigenstate of the creation operator with eigenvalue 1, implying that
for any normal-ordered polynomial of the ladder operators one has
\begin{equation}
\left< S \right| Q(\{a^\dag_i\},\{a_i\})=\left< S \right| Q(\{1\},\{a_i\}).
\end{equation}
Thus we can write expectation value Eq.(\ref{expectation}) as
\begin{equation}
\left< O(t) \right>= \left< S \right| O(\{a_i\}) \left| \Phi (t) \right>.
\end{equation}
We can write this expectation value as coherent-state path integral:
\begin{equation}
\left< O(T) \right>= \frac{\int \prod_i D\hat{\psi}_iD\psi_i O(\{\psi_i\})e^{-S[\hat{\psi}_i,\psi_i;T]}}
{\int \prod_i D\hat{\psi}_iD\psi_i e^{-S[\hat{\psi}_i,\psi_i;T]}},
\end{equation}
where the action is given by
\begin{equation}
S[\hat{\psi}_i,\psi_i;T]= \sum_i \left( \int_0^T dt \left[
\hat{\psi}_i(t)\frac{\partial}{\partial t}\psi_i(t)+H_i(\{\hat{\psi}_i(t)\},\{\psi_i(t)\})\right]\right).
\end{equation}
Performing the continuum limit:
\begin{eqnarray}
\nonumber
\sum_i \to b^{-1} \int dx, \qquad \psi_i(t) \to b\psi(x,t), \qquad \hat{\psi}_i(t) \to \hat{\psi}(x,t), \\
\sum_{\{e\}} [\psi_e(t)-\psi_i(t)] \to b^3 \frac{\partial^2}{\partial x^2} \psi(x,t),
\end{eqnarray}
we get the action:
\begin{eqnarray}
\nonumber
S[\hat{\psi},\psi;T]= \int dx \left[ \int^T_0 dt \left( \hat{\psi}(x,t) \left[\frac{\partial}{\partial t}-
D \frac{\partial^2}{\partial x^2} \right]\psi(x,t) \right. \right. \\
\left. \left. -\lambda_0 [1-\hat{\psi}(x,t)^2]\psi(x,t)^2
+\sigma[1-\hat{\psi}(x,t)]\hat{\psi}(x,t)\psi(x,t) \right) \right],
\label{action}
\end{eqnarray}
where $\lambda_0=b\lambda$.

\section{perturbation theory}
\label{perturbation}

In order to study perturbatively this field theory we will perform a change of variables to rend the action
Eq.(\ref{action}) dimensionless:
\begin{equation}
t \to \frac{t}{\sigma}, \qquad x \to \sqrt{\frac{D}{\sigma}}x, \qquad \hat{\psi} \to \hat{\psi},
\qquad \psi \to \frac{\sigma}{\lambda}\psi,
\end{equation}
this way we get:
\begin{eqnarray}
\nonumber
S(\hat{\psi},\psi)=\epsilon^{-1} \int dx dt \left( \hat{\psi}(x,t) \left[\frac{\partial}{\partial t}-
\frac{\partial^2}{\partial x^2} \right]\psi(x,t) \right. \\
\left. -[1-\hat{\psi}(x,t)^2]\psi(x,t)^2
+[1-\hat{\psi}(x,t)]\hat{\psi}(x,t)\psi(x,t) \right),
\label{action2}
\end{eqnarray}
where $\epsilon^{-1}=\frac{\sqrt{D\sigma}}{\lambda_0}$. We will use from now on some standard results
involving functionals and functional integrals in field theory, they can be seen for instance
in~\cite{zinn}.
The functional $Z(\hat{\eta},\eta)$ with external sources is:
\begin{equation}
\label{functional}
Z(\hat{\eta},\eta)= \int D\hat{\psi}(x,t) D\psi(x,t)
e^{-\frac{1}{\epsilon} \left(S+\int dx dt [\hat{\eta}(x,t)\psi(x,t)+\hat{\psi}(x,t)\eta(x,t)]\right)}.
\end{equation}
Using the steepest-descent procedure, we know that the functional integral Eq.(\ref{functional}) in the
limit $\epsilon \to 0$ is dominated by the saddle points:
\begin{eqnarray}
\label{saddle1}
\frac{\delta S}{\delta \psi(x,t)}&=&\hat{\eta}(x,t), \\
\label{saddle2}
\frac{\delta S}{\delta \hat{\psi}(x,t)}&=&\eta(x,t).
\end{eqnarray}
Eqs.(\ref{saddle1},\ref{saddle2}) in the absence of the external sources ($\eta=\hat{\eta}=0$)
are the mean-field
equations for the reaction-diffusion process. We will study perturbatively the functional integral
Eq.(\ref{functional}) in a neighborhood of the ``classical field'', say, the solutions of the saddle-point
equations (\ref{saddle1},\ref{saddle2}):$\psi_c,\hat{\psi}_c$. Thus we will use the expansion:
\begin{eqnarray}
\psi=\psi_c+\sqrt{\epsilon}\chi, \\
\hat{\psi}=\hat{\psi}_c+\sqrt{\epsilon}\hat{\chi},
\end{eqnarray}
and expanding the action in powers of $\epsilon$ we find:
\begin{eqnarray}
\nonumber
S(\hat{\psi},\psi)-\eta\hat{\psi}-\hat{\eta}\psi=S(\hat{\psi}_c,\psi_c)-\eta\hat{\psi}_c-\hat{\eta}\psi_c+ \\
\nonumber
\frac{\epsilon}{2} \int dx_1 dx_2 dt_1 dt_2 \left[ \left.
\frac{\delta^2 S}{\delta \psi(x_1,t_1) \delta \psi(x_2,t_2)}
\right|_{\psi=\psi_c} \chi(x_1,t_1) \chi(x_2,t_2) \right. + \\
\nonumber
\left. \frac{\delta^2 S}{\delta \hat{\psi}(x_1,t_1) \delta \hat{\psi}(x_2,t_2)}
\right|_{\hat{\psi}=\hat{\psi}_c}\hat{\chi}(x_1,t_1) \hat{\chi}(x_2,t_2) + \\
\left.
\left. 2\frac{\delta^2 S}{\delta \hat{\psi}(x_1,t_1) \delta \psi(x_2,t_2)}
\right|_{\psi=\psi_c,\hat{\psi}=\hat{\psi}_c}\hat{\chi}(x_1,t_1) \chi(x_2,t_2)\right] + O(\epsilon^{3/2}).
\end{eqnarray}
It is very important to note at this point that it is the edge of the front that leads to the marginal
stability criterium, say, to the velocity selection. And the edge of the front is characterized by a low
occupation number, so we can neglect the terms proportional to $\chi^2$ and $\hat{\chi}^2$, that reflect the
presence of more than one particle at the corresponding site, as we consider this event to be unlikely if
we go far enough in the edge. We can see this clearly if we remind that $\chi$ and $\hat{\chi}$ are the
eigenvalues of the annihilation and creation operators respectively, and this way any of them squared reflects
the possible presence of two particles in the same place.

The functional integral at this order becomes:
\begin{equation}
Z(\hat{\eta},\eta) \sim Z_0(\hat{\eta},\eta) \int D\hat{\chi}D\chi e^{-\int dx_1dx_2dt_1dt_2
\frac{\delta^2 S}{\delta \hat{\psi}_c(x_1,t_1) \delta \psi_c(x_2,t_2)}\hat{\chi}(x_1,t_1)\chi(x_2,t_2)},
\end{equation}
where
\begin{equation}
Z_0(\hat{\eta},\eta)=e^{-\frac{1}{\epsilon}[S(\hat{\psi}_c,\psi_c)-\hat{\eta}\psi_c-\hat{\psi}_c\eta]},
\end{equation}
and therefore:
\begin{equation}
Z(\hat{\eta},\eta)=NZ_0(\hat{\eta},\eta) \left[ \mathrm{det}
\frac{\delta^2 S}{\delta \hat{\psi}_c(x_1,t_1) \delta \psi_c(x_2,t_2)} \right]^{-1}.
\end{equation}
The normalization factor $N$ is fixed by the condition $Z(0,0)=1$.

The connected generating functional $W(\hat{\eta},\eta)=\epsilon \mathrm{ln}Z(\hat{\eta},\eta)$
at this order is then:
\begin{equation}
W(\hat{\eta},\eta)=W_0(\hat{\eta},\eta)+\epsilon W_1(\hat{\eta},\eta) + O(\epsilon^2),
\end{equation}
where
\begin{equation}
W_1(\hat{\eta},\eta)=-\left[ \mathrm{tr \hspace{0.1cm} ln} \left.
\frac{\delta^2 S}{\delta \hat{\psi}_c(x_1,t_1) \delta \psi_c(x_2,t_2)} \right|_{\hat{\eta},\eta}-
\mathrm{tr \hspace{0.1cm} ln} \left.
\frac{\delta^2 S}{\delta \hat{\psi}_c(x_1,t_1) \delta \psi_c(x_2,t_2)} \right|_{\hat{\eta}=\eta=0}
\right].
\end{equation}
Let us now perform the Legendre transformation
\begin{equation}
\Gamma(\hat{\phi},\phi)=\int dx dt [ \eta(x,t)\hat{\phi}(x,t)+\hat{\eta}(x,t)\phi(x,t)-
W_0(\hat{\eta},\eta)-\epsilon W_1(\hat{\eta},\eta)]+O(\epsilon^2).
\end{equation}
At one-loop order the 1PI functional is:
\begin{equation}
\Gamma(\hat{\phi},\phi)=S(\hat{\phi},\phi)+\epsilon \Gamma_1(\hat{\phi},\phi)+O(\epsilon^2),
\end{equation}
with:
\begin{equation}
\Gamma_1(\hat{\phi},\phi)=\mathrm{tr} \left[\mathrm{ln} \frac{\delta^2 S}
{\delta \hat{\phi}(x_1,t_1) \delta \phi(x_2,t_2)}-
\mathrm{ln} \left. \frac{\delta^2 S}{\delta \hat{\phi} \delta \phi} \right|_{\hat{\phi}=\phi=0} \right].
\end{equation}
We can interpret the 1PI functional as an effective action that will lead us to new effective equations of
motion:
\begin{eqnarray}
\label{motion1}
\frac{\delta \Gamma}{\delta \phi}&=&0, \\
\label{motion2}
\frac{\delta \Gamma}{\delta \hat{\phi}}&=&0.
\end{eqnarray}
In our particular case, action (\ref{action2}) gives:
\begin{equation}
\frac{\delta^2 S}{\delta \hat{\phi}\delta \phi}=(\partial_t-\partial_{xx}+1)\delta_2 +
(4\hat{\phi}\phi-2\hat{\phi})\delta_2,
\end{equation}
where $\delta_2=\delta(x_1-x_2)\delta(t_1-t_2)$. We conclude therefore:
\begin{eqnarray}
\nonumber
\Gamma_1(\hat{\phi},\phi)= \int dx dt \left<x,t\right|\mathrm{ln}\left[1+
(\partial_t-\partial_{xx}+1)^{-1}(4\hat{\phi}\phi-2\hat{\phi})\right]\left|x,t\right>=
\\
\int dx dt \left<x,t\right|
(\partial_t-\partial_{xx}+1)^{-1}(4\hat{\phi}\phi-2\hat{\phi})\left|x,t\right>,
\end{eqnarray}
where we have made use of the low occupation number approximation for the front edge and a power series
expansion of the logarithm. If we evaluate the matrix element we get:
\begin{equation}
\Gamma_1=C \int dx dt (4 \hat{\phi}\phi-2\hat{\phi}),
\end{equation}
where the constant $C$ is given by:
\begin{eqnarray}
\nonumber
C=\frac{1}{(2\pi)^2}\int dp \int dw \frac{1}{-iw+p^2+1}=\frac{1}{(2\pi)^2} \int dp \int dw \int dt \Theta(t)
e^{-(p^2+1)t}e^{iwt}= \\
\nonumber
\frac{1}{(2\pi)^2}\int dp \int dt \Theta(t)e^{-(p^2+1)t}\int dw e^{iwt} \\
=\frac{1}{(2\pi)^2}\int dp \int dt \Theta(t)
e^{-(p^2+1)t}2\pi \delta(t)=\frac{1}{4\pi}\int dp = \frac{1}{2b_0},
\end{eqnarray}
where we have used the fact that $\Theta(0)=1/2$ (this property of the Heaviside $\Theta$ function can be
found more rigourosly proven in~\cite{zinn})
and that the integral over the whole momentum space is the
volume of the first Brillouin zone, where $b_0$ is the dimensionless lattice spacing.
Thus the effective action reads:
\begin{equation}
\Gamma(\hat{\phi},\phi)=S(\hat{\phi},\phi)+\epsilon \frac{1}{2b_0}
\int dx dt (4 \hat{\phi}\phi-2\hat{\phi})+O(\epsilon^2),
\end{equation}
and the new equations of motion Eqs.(\ref{motion1},\ref{motion2}) are:
\begin{eqnarray}
\label{eff1}
(\partial_t-\partial_{xx})\phi+2\hat{\phi}\phi^2-2\hat{\phi}\phi+\phi+
\epsilon \frac{1}{2b_0}(4\phi-2)&=&0, \\
\label{eff2}
-(\partial_t+\partial_{xx})\hat{\phi}-2(1-\hat{\phi}^2)\phi+(1-\hat{\phi})\hat{\phi}+
\epsilon \frac{1}{2b_0}(4\hat{\phi})&=&0.
\end{eqnarray}
If we remind that the field $\phi$ represents the expected value of the front density, and that we are
everywhere supposing that we are on the front edge, this quantity should be small enough to consider
$\epsilon \phi$ neligible. This way we get that $\hat{\phi}=1+2\epsilon/b_0$ solves equation Eq.(\ref{eff2}).
Substituting this result in Eq.(\ref{eff1}) and taking into account that $b_0=\sqrt{\sigma/D}b$ we
get
\begin{equation}
\label{effective}
\partial_t \phi= \partial_{xx} \phi +\phi - 2\phi^2 + \frac{\lambda}{\sigma}.
\end{equation}
Note that in this case, contrary to the mean field approach, a positive phase propagates into an
(infinitesimaly) negative phase, something that is clearly unphysical. This is the deterministic expression
of the compact support property of the front, i.e., the front becomes identically zero at a finite value of
$x$. Actually, this property has been rigourosly proven for this kind of fronts~\cite{mueller}.
In this regime, since there are no particles, no reaction is possible, only diffusion from adjacent
sites is allowed, leading to the action:
\begin{equation}
S_{diff}=\int dx \int dt [\hat{\phi}(\partial_t-\partial_{xx}) \phi].
\end{equation}
Since this action is quadratic, it is the effective action to any order, and this implies that the effective
equation of motion for the front propagation at first order in $\lambda/\sigma$ is
\begin{equation}
\label{cutoff}
\partial_t \phi= \partial_{xx} \phi +\left(\phi - 2\phi^2 + \frac{\lambda}{\sigma}\right)\Theta(\phi).
\end{equation}
This last derivation deserves a further explanation. It may be surprising to the reader that Eq.(\ref{effective})
performs such an unphysical behaviour, but it is actually what one would expect {\it a priori}. Indeed, the analysis
of an effective action commonly yields a shift of the fixed points of the original one, that is what has happened here.
This suggests that the correct physical interpretation of the problem should have been the corresponding to a moving
boundary one. This is, at the beginning, we should have had into account two different actions, one for the space full
of particles and one for the empty space, and study the propagation of the boundary between them. This preserves the
physical meaning all along the derivation. It might be desirable to solve this problem without splitting it into two
parts, something that maybe could be done by using stochastic differential equations~\cite{pechenik,doering}.

\section{Front Propagation and Velocity Selection}
\label{front}

To study how the front propagates let us perform the change of variables $\phi=u-\lambda/\sigma$.
At leading order, Eq.(\ref{cutoff}) becomes:
\begin{equation}
\label{cutoff2}
\partial_t u=\partial_{xx} u + (u-2u^2)\Theta \left(u-\frac{\lambda}{\sigma}\right).
\end{equation}
Clearly, fields $u$ and $\phi$ propagate at the same speed. Eq.(\ref{cutoff2}) was heuristically proposed and
studied by Brunet and Derrida~\cite{brunet},
and we will summarize the main conclusions of their work. We will consider
that for sufficiently long times the front will converge to a stationary shape, this is,
$u(x,t)=u(x-ct)=u(z)$, and Eq.(\ref{cutoff2}) becomes
\begin{equation}
\label{stationary}
u''+cu'+(u-2u^2)\Theta\left(u-\frac{\lambda}{\sigma} \right)=0,
\end{equation}
where $c$ is the dimensionless front speed. We can distinguish between three regions in the front edge,
in the first one, the equation of motion is given by
\begin{equation}
u''+cu'+u-2u^2=0,
\end{equation}
in the second one, the field
is small enough that we can linearize to get:
\begin{equation}
u''+cu'+u=0,
\end{equation}
and in the third one $u<\lambda/\sigma$, so we get
\begin{equation}
u''+cu'=0.
\end{equation}
We impose as boundary conditions the continuity of the first derivative between the boundaries of the three
different regions. It can be shown~\cite{brunet} that this leads to a dimensionless velocity
\begin{equation}
c=2-\frac{\pi^2}{\mathrm{ln}^2(\lambda/\sigma)}
\end{equation}
at first order in $\lambda/\sigma$. The front velocity with the corresponding dimensions is thus:
\begin{equation}
v=\sqrt{D\sigma}\left(2-\frac{\pi^2}{\mathrm{ln}^2(\lambda/\sigma)}\right).
\end{equation}
This correction has already been confirmed in numerical simulations~\cite{brunet,pechenik,doering}, and
it shows a very slow convergence to the mean-field velocity in the limit $\lambda/\sigma \to 0$. This shows
that the discretness of the reaction process strongly shifts the velocity to a slower one.

As a final remark, we would like to underline that the cutoff derived is not a particularity of the low dimensional
topology of the problem. Indeed, if one considers the $d$-dimensional problem and performs all the calculations shown
here for the particular case $d=1$, one arrives at the dimensionless equation:
\begin{equation}
\partial_t \phi= \nabla^2 \phi +\left(\phi - 2\phi^2 + \frac{\lambda}{\sigma}\right)\Theta(\phi),
\end{equation}
locally describing the edge of the front. The reason of this independence between the appearance of the cutoff in the
reaction term and the dimensionality of the system is due to the physical origin of the cutoff. It appears as a
consequence of the physical fact that far enough in the right spatial direction (the direction of propagation of
the front) there must be no particles. This is, of course, totally independent of the spatial dimension of the system.
However, the effect of the cutoff on the dynamics of the front does strongly depends on the dimensionality. In one
dimension, we have observed a strong shift on the velocity of the front, while in two dimensions the effect is even
stronger and the presence of the cutoff is the only responsible for the formation of diffusive
instabilities~\cite{kessler}. It would be very interesting to analize the effects of the cutoff in dimensions above
$d=2$, to see if new phenomenology develops or contrary there is a return to the mean field.

\section{Conclusions}
\label{conclusions}
In this work, we have derived from first principles a field theory of reaction-diffusion particles, and we
have used it to study reaction-diffusion propagating fronts. This kind of fronts has been traditionally
studied using deterministic reaction-diffusion equations, like the Fisher equation, that consider an infinite
number of particles. We performed a perturbation expansion in the ratio between the annihilation and the
birth rates, that separates the mean field regime from the real discrete process, and studied the first order
corrections to the equation of motion. A cutoff in the reaction term, a mechanism that has already been
heuristically proposed, appeared in a natural way within our formalism, and leaded us to the velocity
corrections already found in numerical simulations.

Our first-principles analitically derived theory also allowed us to understand the fundamental reasons that
lead to the velocity shift. It is the compact support property of the front, i.e., the fact that the field
is identically zero far enough in the spatial axis what causes such a dramatic effect in a pulled front like
ours.

It is also interesting to compare this work to a former one in the same direction~\cite{levine} which tries to
derive a cutoff in the reaction term, albeit for a different system. The main difference between both works is,
under our point of view, that while this article concerns a system in the continuum space, the other deals
with a lattice. This difference becomes fundamental since the calculations were performed using Stratonovich stochastic
calculus, valid in the lattice, but ill-posed in a continuum space~\cite{rocco}.

Many questions are still to be answered. Different reaction shemes are to be explored, also, the opposite
limit (the annihilation rate large compared to the birth rate) is only conjectured~\cite{doering},
but not analitically
found. Of course, higher dimensionality of the front is a very interesting problem, where new phenomenology
does appear, as shown by Kessler and Levine~\cite{kessler}.
We hope that this and former works will encourage the reader
to attempt to solve these and different problems that appear in this subject.

\begin{acknowledgments}

This work has been partially supported by the Ministerio de Educaci\'{o}n y Cultura (Spain) through Grant No. AP2001-2598 and
by the Ministerio de Ciencia y Tecnolog\'{\i}a (Spain) through Project No. BFM2001-0291.

\end{acknowledgments}

\end{document}